\pdfoutput=1

\documentclass[10pt]{article}

\usepackage[utf8]{inputenc}
\usepackage{amsmath, amssymb}
\usepackage{graphicx}
\usepackage{booktabs}
\usepackage[numbers]{natbib}
\usepackage{authblk}
\usepackage{hyperref}

\usepackage{geometry}
\usepackage{enumitem}
\usepackage{xspace}
\usepackage{ulem}
\usepackage{tabularx}

\newcommand{\ModelingSubsystem}{{Modeling}\xspace}
\newcommand{\EvolutionSubsystem}{{Evolution}\xspace}

\newcommand{\KMConstructor}{\textit{Model Constructor}\xspace}
\newcommand{\KMEvolver}{\textit{Model Updater}\xspace}
\newcommand{\KMAnalyer}{\textit{Model Analyzer}\xspace}

\newcommand{\KMExecutionDriver}{\textit{KM Execution Driver}\xspace}
\newcommand{\KMExecutor}{\textit{KM Executor}\xspace}
\newcommand{\KMValidator}{\textit{Runtime Validator}\xspace}

\newcommand{\KMEAdvisor}{\textit{Evolution Advisor}\xspace}
\newcommand{\KMReasoner}{\textit{Symbolic Reasoner}\xspace}

 \usepackage[table]{xcolor}
 \usepackage{makecell}
 
\title{Agentic Generation and Evolution of Knowledge Models}

\author[1]{Man Zhang}
\author[1]{Tao Yue\thanks{Corresponding author}}
\author[2,5,6]{Nazareno M. Aguirre}
\author[3,5,6]{Diego Garbervetsky}
\author[3,4,6]{Sebastian Uchitel}

\affil[1]{Beihang University, China}
\affil[2]{Universidad Nacional de Río Cuarto, Argentina}
\affil[3]{University of Buenos Aires, Argentina}
\affil[4]{Imperial College London, United Kingdom}
\affil[5]{Guangdong Technion-Israel Institute of Technology, China}
\affil[6]{CONICET, Argentina}

\date{}

\begin{document}

\maketitle

\begin{abstract}
Complex software systems such as autonomous vehicles, robotics increasingly interact with dynamic physical, cyber, and social environments. Reasoning about their behavior, maintaining them under continuous change, and evolving them safely require trustworthy knowledge about the system, its assumptions, and its operating context. Knowledge models (KMs) provide a practical basis for such reasoning, but they may themselves become incomplete, inconsistent, or outdated as systems evolve.
This paper presents \textit{TrustModel}, a vision for the agentic generation and evolution of living KMs. TrustModel comprises three agentic subsystems: \textit{Modeling}, for constructing and updating KMs; \textit{Conformance}, for assessing their alignment with the system and its environment; and \textit{Evolution}, for generating guidance to keep KMs synchronized with emerging changes. We demonstrate how TrustModel can be instantiated for model-based testing and discuss its potential for supporting other MDE activities, such as requirements and assumption monitoring, architectural drift tracking, and change impact assessment. Overall, \textit{TrustModel} positions living KMs as a foundation for dependable engineering of continuously evolving software systems.
\end{abstract}

\section{Introduction}
Complex software systems, such as those found in autonomous vehicles, robotics, cyber-physical systems, and other safety-critical or dynamically evolving domains, increasingly interact with physical environments, external services, human operators, and other systems~\cite{casadei2025software}. As these systems evolve to accommodate new requirements, changing operating conditions, architectural modifications, and emerging functionalities, reasoning about their behavior and maintaining them safely become increasingly difficult~\cite{yue2023ASE, controlsoftware2025}.

Modeling provides an essential means to manage this complexity~\cite{france2007model}. By abstracting from low-level implementation details, models make relevant system knowledge explicit and provide a structured basis for reasoning, analysis, validation, maintenance, and evolution~\cite{burgueno2025automation}. However, the engineering knowledge captured by such models is often distributed across heterogeneous artifacts and may become incomplete, inconsistent, or outdated as the system and its environment change~\cite{michael2025model}. As a result, model-driven engineering (MDE) activities that rely on such knowledge become less reliable when the underlying knowledge models (KMs) no longer reflect the actual system, its engineering artifacts, or its operating environment. This motivates the need for mechanisms that can continuously construct, validate, and evolve KMs so that they remain synchronized with the evolving system and its context, to better empowering MDE activities such as early verification and validation of system behaviors~\cite{cederbladh2025road}.

\begin{figure}[htbp]
    \centering
    \includegraphics[width=0.5\textwidth]{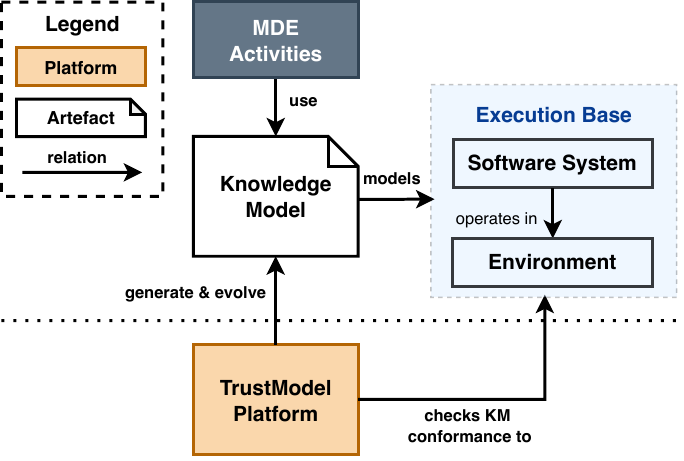}
    \caption{Overall Context of TrustModel. \textit{TrustModel Platform} generates and evolves a KM for \textit{MDE Activities} by checking its conformance against \textit{Execution Base}, namely the running software system and its operating environment.}
    \label{fig:context}
\end{figure}

KMs provide explicit and abstract representations of system knowledge, including structural elements, behavioral specifications, requirements, assumptions, architectural intent, dependencies, constraints, and traceability relationships. Different types of KMs can support a range of MDE activities that contribute to system dependability and long-term maintenance, such as model-based testing (MBT), requirements and assumption monitoring, architectural drift tracking, and change impact assessment~\cite{LLM4MDESurvey}. The top part of Figure~\ref{fig:context} illustrates how KMs serve as foundational assets for these activities. It also highlights that, to remain useful, KMs cannot be treated as static artifacts; they need to be maintained as living models that continuously and adequately reflect the evolving software system, its engineering artifacts, and its operating environment.


Constructing KMs to support MDE activities has long been a central concern in the MDE community. Existing work has investigated how models can be constructed and used as primary engineering artifacts for activities such as MBT~\cite{camilli2021uncertainty,turker2024accelerating}.
Beyond their initial construction, keeping KMs ``alive'', i.e., continuously aligned with the evolving software system and its operating environment, is equally important. Once models fall out of sync, the reasoning and engineering activities they support become unreliable, making failures increasingly difficult to anticipate~\cite{araujo2023testing}. Related efforts have addressed model evolution, uncertainty-aware model maintenance, and safety-oriented model adaptation~\cite{hebig2017approaches,zhang2017uncertainty,tan2025safety}. However, these approaches remain insufficient for maintaining KMs as living models throughout the lifetime of modern software systems, especially when KM evolution must be continuously guided by changes observed in both the system and its operating environment~\cite{yue2023ASE}.

To address the challenge of keeping KMs conformant with the system and its operating environment, this paper proposes an agent-based platform, that we call \textit{TrustModel}. Our approach treats model evolution as a primary concern, together with corresponding agentic subsystems, that collaboratively keep KMs alive:
\begin{itemize}
    
    \item \textit{Modeling}, driven by an agentic subsystem that supports both the initial generation and subsequent update of KMs. The agentic subsystem uses system requirements and the target metamodel for model generation, and additionally considers previous model versions for model update. In both cases, the agentic subsystem can interact with tools for syntactic and semantic analysis to produce syntactically valid and semantically consistent KMs.
    
    \item \textit{Conformance}, driven by an agentic subsystem responsible for assessing the alignment between the KM and the running system together with its operating environment. This subsystem has access to both the system and the KM, proactively seeks evidence of their conformance, and produces conformance feedback to guide subsequent KM evolution.
    
    \item \textit{Evolution}, driven by an agentic subsystem that takes conformance feedback from the conformance subsystem as input and produces concrete evolution guidance for the modeling subsystem to consume. This subsystem reasons over the conformance feedback, domain knowledge, system documentation, and available KM versions to determine how the KMs should be evolved.
\end{itemize}
Together, these three agentic subsystems form the core of \textit{TrustModel}, enabling the continuous, trustworthy evolution of KMs to support dependable MDE activities throughout the system's operational life.
Their design depends on the nature of a KM to be evolved, which in turn depends on the MDE activity that the KM is intended to support. For instance, supporting MBT requires the KM to include behavioral models, whereas supporting architectural drift tracking requires the KM to capture structural constraints. To illustrate how the \textit{TrustModel} platform supports KM evolution, we describe how the \textit{TrustModel} architecture can be instantiated for MBT.

The rest of the paper is structured as follows. Section~\ref{sec:stateOfArt} discusses the literature. Section~\ref{sec:TM architecture} presents the \textit{TrustModel} architecture. Section~\ref{sec:TM4MBT} discusses how the \textit{TrustModel} architecture can be instantiated to support MBT. Section~\ref{sec:discussion} provides a discussion, and Section~\ref{sec:conclusion} concludes the paper.

\section{The Literature}
\label{sec:stateOfArt}
As previously discussed, \textit{TrustModel} is envisioned as a platform organized around three agentic subsystems for KM modeling, conformance assessment, and evolution guidance. These subsystems cannot be realized merely as conventional model management functions; they require agents capable of understanding heterogeneous engineering artifacts, reasoning over modeling knowledge, generating and revising KMs, assessing their conformance, and guiding their evolution. Recent advances in large language models (LLMs), together with their rapidly expanding role in software engineering (SE), make LLMs the principal enabler for implementing these agentic capabilities~\cite{liu2024large}. Therefore, LLM-based agents constitute the main technical foundation for realizing the three agentic subsystems of \textit{TrustModel}. 

In this section, we focus on LLM-based agents in SE, with particular attention to their use in MDE. We especially examine existing work on LLM-supported model generation, model evolution, and their potential to support MBT.

\paragraph{LLMs for SE in General}
Liu et al.~\cite{liu2024large} conducted a systematic literature review (SLR) on LLM-based agents in SE, analyzing 106 primary studies to examine how such agents are designed and applied across various phases of the software development lifecycle, including requirements engineering, code generation, static checking, testing, fault localization, repair, and end-to-end software development and maintenance. Results show that the majority of the primary studies focus on code generation, followed by testing and end-to-end software development. He et al.~\cite{he2025llm} provided a comprehensive review of LLM-based multi-agent systems for SE, with a particular focus on how multiple LLM-driven agents collaborate to solve complex SE tasks. However, there is no concrete proposal for multi-agent-based KM generation and evolution. Both reviews focus on SE in general, rather than specifically targeting the use of LLMs in MDE.

\paragraph{LLMs for Model Generation and Evolution}
Zhang et al.~\cite{LLM4MDESurvey} very recently presented a comprehensive SLR of 228 primary studies to particularly examine how LLMs are applied in MDE, covering their integration into MDE workflows and the state of empirical evaluation. The SLR highlights three key findings. First, the research landscape is rapidly expanding but heavily skewed toward model generation, which accounts for 89 out of 153 studies that propose concrete methods. This trend reflects the longstanding challenge of manual model construction as a barrier to MDE adoption. Second, model generation approaches are diverse and heterogeneous, spanning a wide range of categories, from standard modeling languages and DSL development to ontology learning and enterprise modeling, which suggests broad applicability but a lack of methodological consolidation. Third, \textbf{model evolution remains largely underexplored}, with only two studies~\cite{kong2025collaboration, zhang2025leveraging} addressing it, pointing to a significant gap in supporting the continuous adaptation and co-evolution of models in dynamic environments. 

As reported by Zhang et al.~\cite{LLM4MDESurvey}, the work by Kong et al.~\cite{kong2025collaboration} is the only study that addresses both model generation and model evolution. It first uses ChatGPT to generate two key artifacts from the original system requirements: (i) a Requirements Traceability Matrix (RTM), capturing relationships across requirements and their links to other artifacts, and (ii) a SysML-based Generative Requirements Structure Model (GRSM), representing requirement structures and dependencies. Model evolution is then carried out through a structured, model-based process: the RTM and GRSM are updated according to new requirements, and their previous and revised versions are compared to identify change impact and scope. \textbf{This evolution is therefore process-driven and model-based}, with the \textbf{LLM assisting in artifact generation and updates}, \textbf{rather than directly performing autonomous model evolution}. 

In the literature, existing tools and platforms, such as Papyrus Moka~\cite{papyrus_moka} and MagicDraw Cameo Simulation Toolkit~\cite{cameo_simulation_toolkit}, provide support for model execution and validation. At the same time, integrating LLMs with external tools has become a widely accepted paradigm for building agentic systems, where LLMs act as reasoning and orchestration components while tools provide specific capabilities, access to structured artifacts, execution mechanisms, and feedback channels. This paradigm is already reflected in LLM-based MDE: according to the survey~\cite{LLM4MDESurvey}, current LLM-based MDE solutions have been integrated with a variety of MDE tools and platforms, with PlantUML and the Eclipse Modeling Framework (EMF) being the most commonly used. However, these integrations primarily focus on model generation and manipulation, rather than model execution or simulation. To the best of our knowledge, no existing work tightly integrates LLM-based MDE approaches with KM execution or simulation environments as we do. Although two studies~\cite{von2024toward, sugawara2025extracting} report integration with Cameo Systems Modeler, they do not leverage its simulation capabilities, and thus stop short of enabling execution-driven validation and feedback.

\paragraph{LLM-based Model-based Testing (MBT)}
From the SLR~\cite{LLM4MDESurvey}, we also identified eight studies that are about \textbf{LLM-based MBT}, covering model-based test generation, test execution, test incident reporting, etc., as summarized in Table~\ref{tab:MBT}. Specifically, most MBT approaches leverage LLMs to transform heterogeneous inputs, such as requirements in natural language, domain-specific scenarios, or structured models (e.g., XML), into executable test artifacts, including test scenarios and test cases. Beyond generation, a smaller number of works explore test execution (e.g., UI-level interaction) and model comprehension, such as diagnosing safety violations from scenario descriptions. Only limited efforts address more integrated pipelines that combine constraint generation, model construction, and test derivation, indicating that end-to-end support remains underexplored.

\begin{table}[htbp]
\caption{Summary of LLM-based MBT studies collected from~\cite{LLM4MDESurvey}}
\label{tab:MBT}
\centering
\scriptsize
\setlength{\tabcolsep}{3pt}
\renewcommand{\arraystretch}{1.15}

\begin{tabular}{p{0.03\textwidth} p{0.16\textwidth} p{0.22\textwidth} p{0.18\textwidth} p{0.24\textwidth} p{0.08\textwidth}}
\toprule
\textbf{Ref.} & \textbf{Task} & \textbf{LLM Input} & \textbf{LLM Output} & \textbf{KM} & \textbf{Domain} \\
\midrule

\cite{deng2025target}
& Model-based test gen.
& Traffic rules (NL)
& Test scenarios (DSL)
& DSL
& ADS \\

\cite{azimi2025model}
& Model-based test gen. \& exec.
& Test scenarios (NL)
& UI models (nd-FSM)
& nd-FSM, behavioral
& Smart TVs \\

\cite{zheng2024testing}
& Model-based test gen.
& Scenario description (DSL)
& Scenario representations, unspecified
& DSL
& ADS \\

\cite{lu2024diavio}
& Model comprehension
& Scenario descriptions (DSL)
& Diagnostic results (NL)
& DSL, behavioral
& ADS \\

\cite{naimi2024new}
& Model-based test gen.
& Models (XML)
& Test cases (NL)
& UML/use case model, behavioral
& General \\

\cite{petrovic2024llm}
& Model-based test gen.
& Requirements (NL)
& Test scenarios (JSON)
& DSL, structural
& ADS \\

\cite{naimi2024automating1}
& Model-based test gen.
& Models (XML)
& Test cases (NL)
& UML/class model, structural
& General \\

\cite{chenail2025test}
& Model-based test gen.
& Use case specifications (NL)
& Test specification (DSL)
& DSL/use case model, behavioral
& IoT \\

\bottomrule
\end{tabular}

\vspace{1mm}
\footnotesize
\noindent\textbf{Abbreviations:} NL: Natural Language; DSL: Domain-Specific Language; XML: eXtensible Markup Language; JSON: JavaScript Object Notation; UML: Unified Modeling Language; nd-FSM: non-deterministic Finite State Machine; ADS: Autonomous Driving Systems; IoT: Internet of Things; gen.: generation; exec.: execution.

\end{table}

Across these studies, \textbf{LLMs primarily function as translation and synthesis mechanisms} between different abstraction levels and representation formats, \textbf{rather than supporting continuous model evolution or closed-loop engineering processes}. In terms of modeling practices, DSLs are the dominant medium, particularly in scenario-driven settings, while traditional MDE technologies such as UML and Ecore/OCL appear less frequently. Generated or manipulated models are largely behavioral, reflecting the central role of scenarios in MBT workflows.
From an application perspective, the literature is heavily skewed toward ADS, with additional but less frequent use cases in general software systems, IoT, and UI-based platforms. This distribution suggests that current LLM-based MBT research is most mature in safety-critical, scenario-intensive domains, while \textbf{its applicability to broader MDE contexts remains limited}.

Although these works demonstrate the potential of LLMs in automating key MBT tasks, \textbf{the field remains relatively immature}. In particular, there is a notable \textbf{lack of research on model evolution} driven by test execution feedback. Existing approaches largely treat test models as static artifacts generated upfront, without explicitly addressing how execution results (e.g., failures, coverage gaps, or runtime observations) can be used to iteratively refine, adapt, or evolve the underlying models. This absence of feedback-driven, closed-loop MBT highlights a critical gap and an important direction for future research.

\section{The TrustModel Architecture}\label{sec:TM architecture}

\begin{figure}
    \centering
    \includegraphics[width=0.85\linewidth]{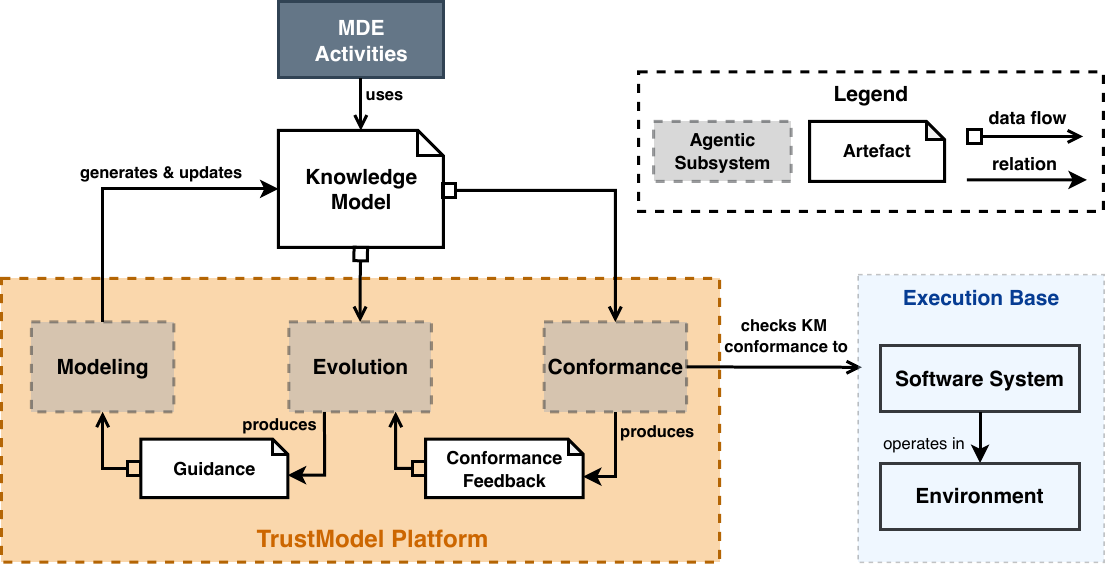}
    \caption{TrustModel Architecture. The \textit{TrustModel Platform} is organized around three agentic subsystems: reactive \textit{Modeling}, proactive and reactive \textit{Evolution}, and proactive and reactive \textit{Conformance}, which are linked through \textit{Guidance} and \textit{Conformance Feedback} while interacting with the \textit{Execution Base}.}
    \label{fig:abstractvision}
\end{figure}

The TrustModel platform aims to address the challenge of supporting the construction and continuous evolution of KMs by means of an agentic architecture. By agent we mean an LLM autonomously using tools in a loop, without a strict predefined workflow, to achieve a goal~\cite{AnthropicAgent1,AnthropicAgent2}. 
Hence, as shown in Figure~\ref{fig:abstractvision}, we propose an architecture, which is composed of the following three agentic subsystems: \textit{Modeling}, \textit{Conformance}, and \textit{Evolution}. 

\paragraph{Modeling agentic subsystem}

The modeling agentic subsystem is responsible for all modeling activities within the platform. In particular, it constructs initial KMs from requirements and documentation, and it also deals with the updating of existing KMs when changes are needed. Its operation is triggered by received requirements and documentation, when models are initially constructed, as well as by strategic guidance received from the evolution subsystem. Upon receiving a guidance, the modeling subsystem reactively constructs and modifies a KM accordingly. 

Regardless of whether a KM is built from scratch or evolved from an existing one, the resulting model must be syntactically correct (i.e., a valid instance of the KM metamodel) and semantically consistent, and thus the agent interacts with tools that help it guarantee these model characteristics. The agentic subsystem works by combining various tools. It uses an LLM-based approach to generate candidate KMs that can be iteratively refined after using a syntactic KM analyzer and consistency model checking tools for instance. To resolve semantic inconsistencies, the agentic subsystem may also have to resort to external domain knowledge. For instance, an inconsistency between a KM (e.g., state machine) and a formally specified temporal constraint will require changing one or the other based on an understanding of the context.

\paragraph{Conformance agentic subsystem}

The goal of the conformance agentic subsystem is to produce evidence of whether a KM conforms, or fails to conform, to the actual system and its environment.
The nature of conformance checking is highly coupled with the nature of the KM itself. If the KM includes behavioral information, conformance can be checked by actively exercising the software system in a testing environment and comparing the observed behavior against the KM, or by monitoring the system in production and then performing a comparison. If the KM includes structural information about the system, conformance checking may require, for example, static analysis of the source code or inspection of deployment configuration files.

Consequently, tools required by the conformance subsystem to achieve its goal may vary significantly. Some tools may be off-the-shelf, such as testing and instrumentation frameworks, logging infrastructure, coverage reporters, assertion checkers, and trace comparators. Others may be bespoke components, which could themselves include LLMs.

Note that the conformance subsystem operates both proactively and reactively. It may react to changes in an KM or the software system by initiating checks that focus specifically on the changes made (e.g., through targeted fuzzing). At the same time, it may act proactively, since changes in the environment can yield insights and behaviors worth documenting in the KM.

\paragraph{Evolution agentic subsystem}

The goal of the evolution subsystem is to provide guidance on how to evolve a KM based on positive and negative conformance feedback produced by the conformance subsystem. 
%
A key strategic decision this subsystem must make when it detects a conformance issue is whether the software system or the KM should be changed, or both. Another strategic decision concerns the kind of changes to apply to the KM. Examples of KM evolution strategies include adding a new requirement that better captures the current behavior of the software system, updating assumptions about the environment that have been monitored and found invalid, or introducing new model elements that had not previously been included in the scope of the KM.

Developing an evolution strategy requires integrating a significant amount of contextual information to synthesize a meaningful suggestion. The decision of whether a non-correspondence indicates a software bug or a problem in the KM is not necessarily clear-cut; in a manual setting, it would likely require input from a domain expert. To address this, we exploit the probabilistic inference of LLM-based agents.
Thus, the evolution subsystem develops strategies using a variety of tools. It may access databases containing domain knowledge (e.g., regulations, standards, requirements documents, user manuals), retrieve logs and historical data from the software system, or even consult the software code repository for commit information. The agentic subsystem may also employ symbolic reasoning tools to reduce the likelihood of issuing incorrect recommendations or to inform decision-making with sound, inferred knowledge.

Like the conformance subsystem, the evolution subsystem works both reactively and proactively. It reacts to evidence of non-conformance, but may also proactively search, for instance, the internet to discover new regulations or emerging trends in user comments that are worth considering when suggesting KM evolutions.

\paragraph{Summary}
Table~\ref{tab:trustmodel-agents} summarizes the three agentic subsystems, their goals, operation modes and representative tools that they may be invoked within their control loops for perception, reasoning, tool use, and goal-achievement evaluation. 

\begin{table*}[ht]
\footnotesize
\centering
\centering
\caption{Summary of the Agentic Subsystems of the TrustModel Platform}
\label{tab:trustmodel-agents}
\renewcommand{\arraystretch}{1.4}
\begin{tabular}{|p{1.6cm}|p{5cm}|p{1cm}|p{4.9cm}|}
\hline
\textbf{Agentic Subsystem} & \textbf{Goal} & \textbf{Mode} & \textbf{Examples of Tools} \\
\hline
Modeling
&
Construct a syntactically correct and semantically consistent KM or update an existing one based on an evolution strategy.
&
Reactive
&
Automated formalizer,
Syntactic analyzer,
Semantic consistency checker,
Domain knowledge retriever. 
\\
\hline
Conformance
&
Produce evidence of conformance (or non-conformance) between the KM and the software system and its environment.
&
Proactive and reactive
&
Testing and instrumentation frameworks, logging infrastructure, coverage reporters, assertion checkers, trace comparators, static analyzers, model executors, synthetic data generators.\\
\hline
Evolution 
&
Provide guidance on how to evolve the KM based on conformance feedback received from the conformance subsystem.
&
Proactive and reactive
&
Access to technical standards library, code repositories, and symbolic reasoners (e.g., model checkers, solvers, program analysis tools).
\\
\hline
\end{tabular}
\end{table*}

\section{TrustModel for Model Based Testing}\label{sec:TM4MBT}
In this section, we show how the TrustModel architecture might be instantiated to address the evolution of behavior models (i.e., test models) intended to support MBT. We first describe, through an example, a series of interventions performed by the three agentic subsystems to enable evolution of behaviour models. We then provide details on the instantiated TrustModel architecture for MBT and specifically on how each of its subsystems might be designed.

\subsection{Running Example}

Consider a smart home energy management system (EMS) that integrates automated control with human decision-making to balance occupant comfort, energy efficiency, and user preferences. The system autonomously manages devices such as Heating, Ventilation, and Air Conditioning (HVAC), lighting, and household appliances, while residents retain the ability to intervene by confirming, rejecting, or overriding system decisions. The EMS is required to prioritize comfort when occupants are present, transition to an energy-saving mode when the home is unoccupied, and explicitly engage users whenever automated adjustments may significantly impact comfort or energy cost.

\paragraph{Step \textcircled{1}: Initial Model Construction}

\begin{figure}[t]
    \centering
        \centering
        \includegraphics[width=0.9\textwidth]{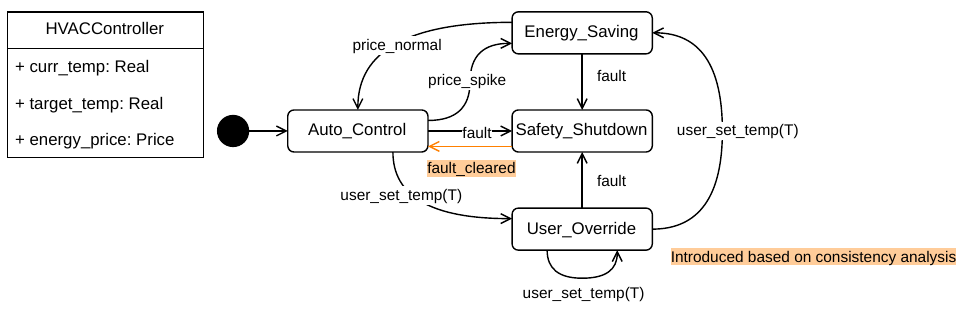}
        \caption{Initial State Machine of EMS's HVAC controller (output of \ModelingSubsystem Subsystem, Step \textcircled{1}). Highlighted elements are introduced based on conformance assessment.}
        \label{fig:init_sm}
\end{figure}

The modeling subsystem takes as input system requirements and documentation for the EMS, identifies relevant information for the construction of the KM, and produces a syntactically valid and semantically consistent KM. Since the goal is to enable MBT, the KM is expressed as a SysML v2 model (with part-based structural views to describe system composition and interfaces, and state-machine-based behavioral views to capture state-dependent execution logic). The model also captures essential requirements such as temperature must be within $[20,26] \,^\circ\text{C}$ unless overridden, and the system must minimize energy cost during peak pricing periods. The obtained state machine is shown in Figure~\ref{fig:init_sm}. Additionally, the subsystem formalizes relevant requirements as signal temporal logic (STL)~\cite{donze2013signal} properties (e.g., it is always the case that, when in autocontrol, the temperature is in the 20-26 degrees range, formally $\textit{FR}_1 =\Box (\textit{Auto\_Control} \implies  20 \leq \textit{curr\_temp} \leq 26)$), to facilitate conformance checking between the KM and the running system. 

\begin{figure}[t]
        \centering
        \includegraphics[width=\textwidth]{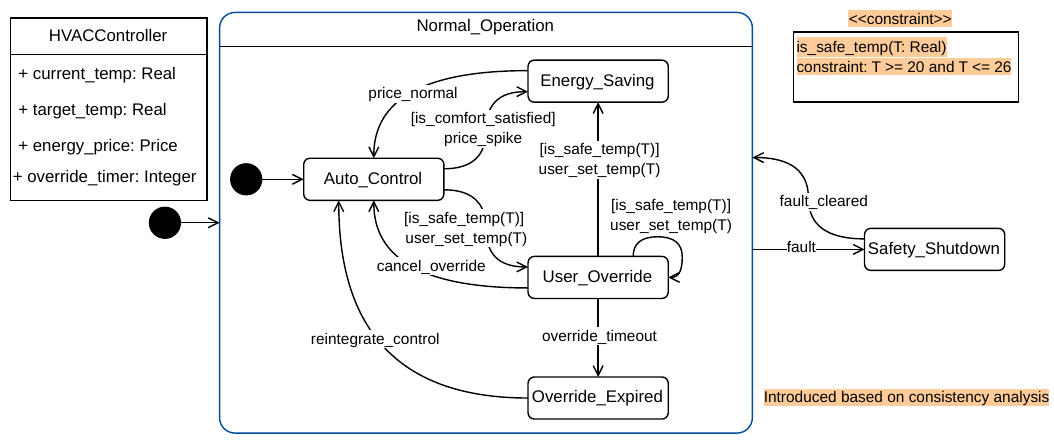}
        \caption{Evolved State machine of EMS's HVAC Controller with Guidance from \KMEAdvisor (Output of \KMEvolver Step \textcircled{3}). Highlighted elements are introduced based on conformance assessment.}
        \label{fig:updated_sm}
\end{figure}

\paragraph{Step \textcircled{2}: Evolution Proposal Based on Domain Knowledge}

The evolution subsystem proactively gathers domain knowledge from external sources, including safety standards, energy regulations, and human-automation interaction guidelines. From this analysis, it synthesizes guidance to refine the KM, such as user overrides expiring after 2 hours ($G_1$), comfort taking priority over cost ($G_2$), and unsafe temperature requests being rejected ($G_3$).

\paragraph{Step \textcircled{3}: Modeling Based on Guidance}

The modeling subsystem incorporates guidance $G_1$--$G_3$ by introducing new model elements (e.g., an \textit{override\_timer} variable and an \textit{Override\_Expired} state), guarded conditions on transitions (e.g., ignoring user override requests if selected temperature is unsafe), and structural refinements. The obtained evolved state machine is shown in Figure~\ref{fig:updated_sm}.

\paragraph{Step \textcircled{4}: Conformance Violation}

The conformance subsystem employs the KM to generate a battery of tests, including realistic user requests (e.g., \textit{user\_set\_temp(28)}) and checks conformance of system behaviour against the KM. The subsystem identifies a violation of   the formalized STL property $FR_1$, with the temperature reaching 28${}^\circ$C while the system being in \textit{Auto\_Control}. It generates a violation trace that is reported to the evolution subsystem.

\paragraph{Step \textcircled{5}: Evolution Proposal Based on Invariant Violation}

The evolution strategy agent receives the violation trace and produces a diagnosis of why the violation occurred. From this analysis, the conformance subsystem generates concrete guidance to be communicated to the modeling subsystem: the guard of \textit{cancel\_override} has to be strengthened with the constraint $20 \leq curr\_temp \leq 26$.

\paragraph{Step \textcircled{6}: Modeling Agent Updates}

The modeling subsystem applies the received guidance, which entails additional implicit modifications beyond the explicit guidance: besides strengthening the guard for \textit{cancel\_override}, the modeling subsystem identifies that guards of other related transitions (e.g., \textit{reintegrate\_control}) also need to be strengthened to maintain overall property satisfaction. As a result, an updated state machine that incorporates these improved guards is generated.

\paragraph{Step \textcircled{7}: Conformance Finds Problem}

The conformance subsystem generates another battery of tests and challenging scenarios (e.g., rapid user inputs). Execution reveals delayed override handling and mode oscillations, producing concrete witnessing sequences that are communicated to the Evolution Subsystem.

\paragraph{Step \textcircled{8}: Feedback and Updated Guidance}

The evolution subsystem derives new guidance from the observed mismatches, closing the TrustModel loop. In analyzing the conformance evidence, the evolution subsystem discovers that the system is not implementing mechanisms to avoid user input flooding. To address this, the evolution subsystem generates guidance to improve the KM with constraints on the frequency of user inputs. Additionally, the subsystem generates a system request for change, indicating that a system update incorporating a debouncing mechanism is necessary to fully align implementation with the evolved KM. This guidance and the corresponding development issue enable continuous KM evolution, progressively improving the safety and trustworthiness of the smart home EMS.

\subsection{Architecture}
In this section, we describe an instantiated version of the TrustModel architecture that would implement the running example described above. 

\subsubsection{Overview}
\begin{figure}[ht]
    \centering
    \includegraphics[width=1\linewidth]{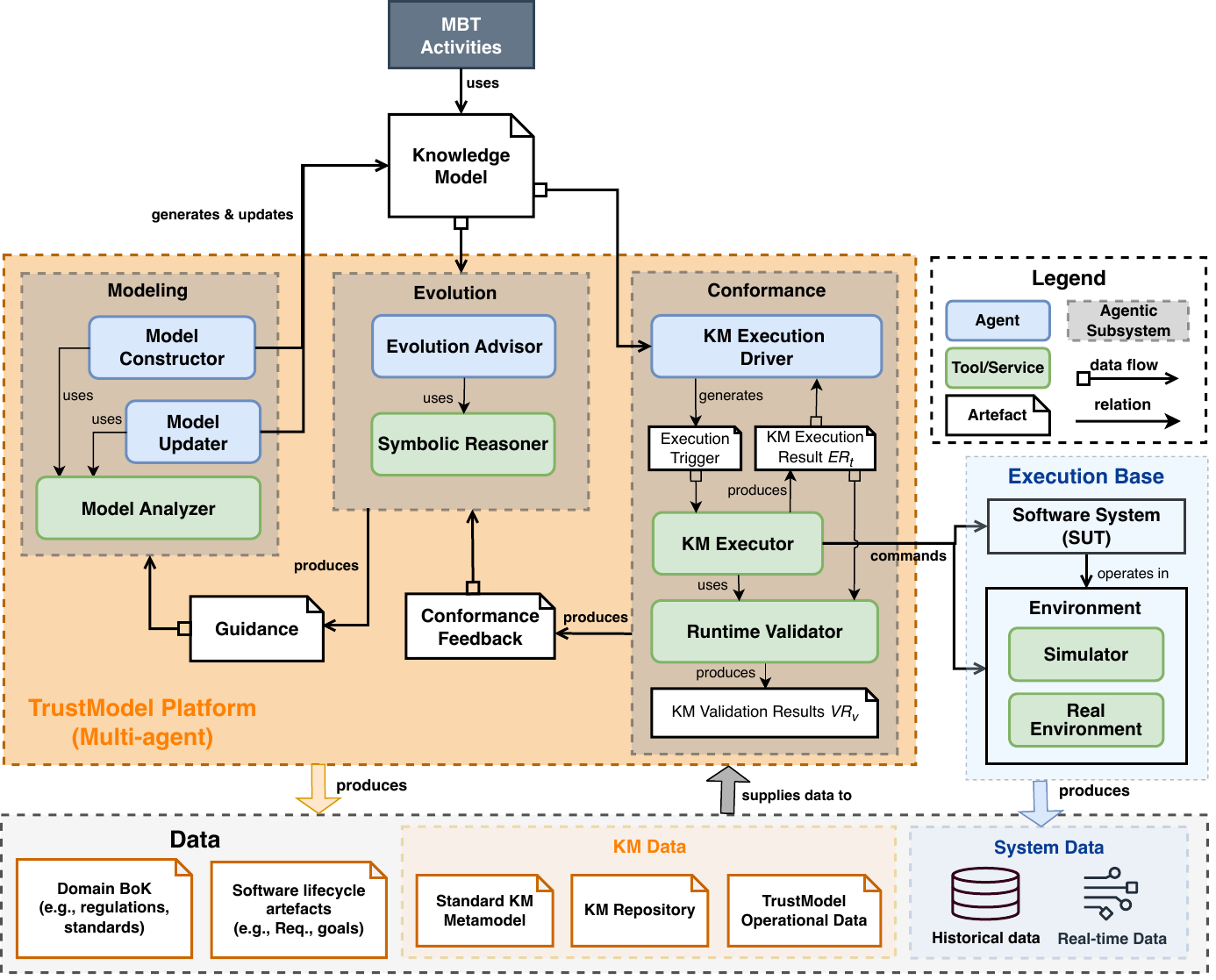}
    \caption{Architecture of \textit{TrustModel} Platform for MBT. The platform instantiates the overall \textit{TrustModel} architecture through three agentic subsystems. Each subsystem is further detailed in terms of its constituent agents and the associated tools and platforms that support their operation.
}
    \label{fig:concretevision}
\end{figure}

The instantiation is depicted in Figure~\ref{fig:concretevision} and includes a refined version of the three agentic subsystems shown in Figure~\ref{fig:abstractvision}. We describe this refinement further down (Subsections \ref{sec:ModelingSubsystem}--\ref{sec:EvolutionSubsystem}). The TrustModel architecture for MBT also incorporates a simulator in the \textit{Execution Base} and also the \textit{Data} component consisting of \textit{System Data}, domain body of knowledge (BoK), software lifecycle artefacts, and \textit{KM Data}. 

\textit{KM Data} comprises three elements: \textit{TrustModel Operational Data}, \textit{KM Repository}, and \textit{Standard KM Metamodel}. These data are consumed by the agents in \textit{TrustModel}. In particular, \textit{Standard KM Metamodel} is used by the \textit{Modeler Constructor} agent to ensure that generated KM elements are in agreement with the expected model structure. More broadly, \textit{KM Data} is continuously maintained to support reuse, agent evolution, and knowledge management. It serves as an internal information hub that accommodates heterogeneous modeling languages and facilitates the efficient construction and evolution of both KMs and agents.

\textit{System Data}, in contrast, is produced by \textit{Execution Base} and includes both historical and real-time data. These data are fed back into the \textit{TrustModel} platform to support KM validation, execution, and evolution. In the smart home scenario, system data may include real-time occupancy data from \texttt{MotionSensor}s, temperature readings from \texttt{SmartThermostat}s, and lighting states from \texttt{LightingControl} devices. This information may then be used by the conformance subsystem to generate execution triggers, by the evolution subsystem to suggest evolution strategies based on the real system usage, and by the modeling subsystem to refine the KM based on observed system behavior, thereby enabling continuous synchronization between the model and the real or simulated system.

\subsubsection{Modeling Agentic Subsystem}
\label{sec:ModelingSubsystem}

The modeling subsystem comprises two agents specialized in generating and updating a KM that conforms to a given KM metamodel. The \textit{Model Constructor} agent is primarily responsible for the initial generation of a KM from system documentation, which may include requirements, design specifications, and implementation details. The agent ensures structural and semantic compliance with the metamodel by leveraging model analysis tools. The \textit{Model Updater} agent, in contrast, focuses on the evolution of an existing KM by incorporating new information and data. It resolves inconsistencies and adjusts the KM to reflect evolving system or domain contexts. Like \textit{Model Constructor}, \textit{Model Updater} also employs model analysis tools (i.e., \textit{Model Analyzer}) to verify syntactic correctness and semantic consistency.

To exemplify how the different components of the subsystem interact, we return to some of the steps described above. 
For step \textcircled{1}, the modeling subsystem takes as input documentation and requirements addressing the system's safety, energy efficiency, and human interaction. Example relevant requirements, identified by the Subsystem, include:
\begin{itemize}
    \item Temperature must remain within $[20,26] \,^\circ\text{C}$ unless overridden by the user; and
    \item The system must minimize energy cost during peak pricing periods.
\end{itemize}
\noindent From these requirements, \KMConstructor generates:
\begin{itemize}
    \item A block with value properties representing the system's current temperature (\textit{curr\_temp}) and the energy price state (\textit{energy\_price});
    \item A state machine with states \textit{Auto\_Control}, \textit{Energy\_Saving}, \textit{User\_Override}, and \textit{Safety\_Shutdown}, as well as events triggered by both environmental changes and human actions (e.g., \textit{price\_spike} and \textit{user\_set\_temp($T$)}).
\end{itemize}

\KMConstructor then calls \KMAnalyer to perform internal syntactic analysis. The subsystem identifies missing value properties (e.g., \textit{target\_temp} for the temperature set by the user, as opposed to the system's current temperature). These are added to the KM by \KMConstructor. Additionally, the subsystem formalizes relevant requirements as STL properties to facilitate conformance checking between the KM and the running system. It produces $\textit{FR}_1 = \Box (\textit{Auto\_Control} \implies 20 \leq \textit{curr\_temp} \leq 26)$, capturing the temperature constraint when in auto-control.

In a second iteration, triggered by \KMConstructor, a semantic analysis performed by \KMAnalyer identifies reachable deadlock states. \KMConstructor processes this feedback to produce a syntactically valid and semantically consistent model, whose state machine is shown in Figure~\ref{fig:init_sm}.

Now let us further describe another step that involved the modeling subsystem. In step \textcircled{3}, the modeling subsystem incorporates guidance G1--G3 by introducing new model elements, guarded conditions, and structural refinements, as outlined in the previous section. However, the \textit{Model Updater} agent may not produce the final evolved state machine (shown in Figure~\ref{fig:updated_sm}) in a single pass.
For example, in a first iteration, the agent generates a version of the state machine in which the \textit{Safety\_Shutdown} state has no outgoing transitions. \KMAnalyer then flags that deadlock states are disallowed according to the KM metamodel. In response, the \textit{Model Updater} agent adds a \textit{fault\_cleared} transition to resolve the deadlock.

Similarly, when incorporating G3 (comfort has priority over cost) and G4 (unsafe temperature requests rejected), the agent introduces guarded conditions using the predicates \textit{is\_comfort\_satisfied} and \textit{is\_temp\_safe(T)}. \KMAnalyer subsequently determines that the KM is not a valid instance of the KM metamodel, because neither predicate is defined. Multiple iterations between \textit{Model Updater} and \KMAnalyer may be required, as \textit{is\_temp\_safe(T)} needs to be defined (see highlighted constraint in Figure~\ref{fig:updated_sm}), and \textit{is\_comfort\_satisfied} may refer to variable \textit{curr\_temp} and constants for upper and lower comfort bounds, that also need to be defined. The agent iteratively adds these missing definitions until the KM becomes syntactically valid and semantically consistent. The resulting evolved state machine is shown in Figure~\ref{fig:updated_sm}.  

\subsubsection{Conformance Agentic Subsystem}
\label{sec:ConformanceSubsystem}

The conformance subsystem is responsible for determining if the KM is an appropriate abstraction of the software system and its environment. 
In the context of MBT, execution is essential for bridging the gap between abstract model specifications and concrete system behavior. 
To support this process, the platform has the \KMExecutionDriver agent that leverages the KM to generate triggers for interacting with the system and its operating environment, while guiding the execution, collecting observed system responses, and comparing them against the expected behaviors to assess conformance.
The actual KM execution is handled by \KMExecutor which takes the triggers and directly interacts with the system and its operating environment, which may be either real or simulated, by sending commands and collecting responses.
Upon execution, \KMExecutor produces results that are subsequently consumed by \KMExecutionDriver to determine the next trigger to generate. This creates a feedback loop that continues until a defined termination criterion is met. The results are also used by \KMValidator for conformance checking.

\textit{KM Execution Driver} can be viewed as playing a role analogous to the test strategy component in MBT. In traditional MBT, test strategies are typically defined over behavioral models (e.g., finite state machines~\cite{turker2024accelerating}, UML state machines~\cite{zhang2019uncertainty}) and are largely coverage-driven, such as state coverage, transition coverage, or path coverage, to systematically derive test cases and triggers for execution. 
\textit{KM Execution Driver} is realized as an agent, enabling a more context-aware and adaptive testing mechanism. Instead of relying solely on fixed coverage criteria, it can leverage the KM, execution history, and broader system context to generate more informed and targeted execution triggers. This aligns with recent LLM-based MBT approaches (as discussed in Section~\ref{sec:stateOfArt}), where LLMs are used to generate test scenarios or test cases from requirements and models. However, existing works primarily focus on test generation, with limited support for intelligent, feedback-driven test strategy adaptation during execution. \KMExecutionDriver extends this paradigm by incorporating reasoning over execution results and system operation context, enabling a more dynamic and closed-loop approach to validation and testing.

In step \textcircled{4}, the conformance subsystem first analyzes the KM to identify relevant constraints and boundaries. It mines constants and value ranges from the formalized properties (e.g., the temperature bounds $[20,26] \,^\circ\text{C}$ from $FR_1$) and from state machine invariants. Based on this analysis, the subsystem performs an input partitioning approach to define specific precondition cases for covering key events—in particular, the \textit{user\_set\_temp($T$)} event. The partitioning yields equivalence classes such as temperatures within the comfort range ($[20,26] \,^\circ\text{C}$), temperatures just above the upper bound ($26$--$28 \,^\circ\text{C}$), and unsafe temperatures ($>28 \,^\circ\text{C}$). From this partitioning, \KMExecutionDriver generates test data that systematically explores each class, eventually leading to violation of the expected behavior. \KMExecutor executes the EMS KMs against the target system via the smart home API. It is \KMValidator that detects that the invariant for the \textit{Auto\_Control} state is violated. From the violation, it produces conformance feedback for the \EvolutionSubsystem.

In step \textcircled{7}, \KMExecutionDriver produces tests that generate \textit{rapid user inputs} to assess debounce control and oscillation prevention.
These triggers drive systematic exploration of human-centered EMS behaviors. In addition, \KMExecutionDriver also generates triggers such as \textit{price\_spike}, covering environmental variations to test the adaptive response of the system to dynamic energy pricing.
\KMExecutor executes the tests and \KMValidator reveals mismatches such as delayed override handling (e.g., \textit{Override latency = 8s}) and mode oscillations.

\subsubsection{Evolution Agentic Subsystem}
\label{sec:EvolutionSubsystem}

The central component of this subsystem is \KMEAdvisor, which is responsible for providing guidance to the modeling subsystem on how to evolve the current KM. This guidance must be inferred from domain knowledge originating from a variety of sources, ranging from system documentation, applicable standards and regulations, and literature covering the system's application domain, to user forums. The guidance may also consider historical information about the KM itself (e.g., from \textit{KM Repository}) and operational data produced by \textit{Execution Base}.

Retrieval is therefore central to \KMEAdvisor. While prior LLM-based multi-agent systems for SE commonly incorporate retrieval mechanisms for knowledge grounding and critic agents for output validation, these roles remain largely decoupled and limited in scope~\cite{he2025llm}. Retrieval mechanisms typically provide passive context, whereas critic agents focus on generic correctness (e.g., syntactic or functional validation) with little consideration of domain-specific semantics. Consequently, existing approaches lack a unified mechanism to enforce domain constraints, interpret execution feedback, and guide model evolution in a coherent manner.

The other important component of this subsystem is \KMReasoner. It is used by \KMEAdvisor to apply formal reasoning techniques that assist in diagnosing the root causes of observed violations, generalizing the problem cause, and evaluating potential repairs. Unlike \KMAnalyer, which focuses on internal model consistency, \KMReasoner concentrates on analyzing discrepancies captured in execution traces, verification results, or constraint violations. It acts as the system's formal diagnostic backbone, helping \KMEAdvisor distinguish between cases where the system incorrectly implements requirements, where the KM inadequately captures domain constraints, or where environmental assumptions are missing or flawed.

\KMReasoner is, in practice, a toolbox that implements a variety of formal techniques, such as unsat core extraction~\cite{10.1007/11814948_5}, causality analysis~\cite{10.1007/s10703-011-0132-2}, interpolation~\cite{10.1007/978-3-540-31980-1_1}, and specification inference~\cite{DBLP:journals/scp/ErnstPGMPTX07,DBLP:conf/icse/TerragniJTP21,DBLP:conf/icse/MolinaPAF21,DBLP:conf/icse/MolinadA22,DBLP:journals/pacmse/NolascoMDGGPUAF24,DBLP:journals/tosem/CasoBGU13}. \KMEAdvisor orchestrates these reasoning tools by first collecting operational data and domain knowledge, then posing formal queries to the reasoning tools for diagnosis. The agent then integrates the diagnosis with its broader domain knowledge to decide whether the KM should be evolved, whether the system implementation requires correction, or whether domain assumptions need revision. As a result, the agent may pose new queries to the symbolic reasoning tools.

In step \textcircled{5}, the conformance subsystem produces evidence, in the form of a system execution trace, that violates the invariant regarding the current system temperature when in \textit{Auto\_Control} ($FR_1$). 
\KMEAdvisor receives the execution log and decides to interact with \KMReasoner for diagnosis generation. Since the execution log contains a violation of the requirements, the STL formalization ($\sigma$) of the log and the STL specification $\phi$ of the state machine in the KM (including the aforementioned STL requirement constraints) are inconsistent; that is, their conjunction is unsatisfiable. \KMReasoner then computes a Craig interpolant for formulas $\sigma$ and $\phi$.

A Craig interpolant of two inconsistent formulas $\sigma$ and $\phi$ is a formula $I$ that predicates over the common vocabulary of $\sigma$ and $\phi$, is consistent with the former, and inconsistent with the latter. It can serve as an explanation of the inconsistency. For our example, the obtained interpolant is the following formula:
$$I = \Box_{[t,t]} (cancel\_override \wedge temp = 28),$$
and relates the violation of requirement $FR_1$ to a specific transition in the logical formalization of the state machine:
$$T_{co} = \Box (cancel\_override \implies Auto\_Control).$$
This interpolant indicates that the reason for the inconsistency stems from the user canceling the override while the temperature is 28 degrees. The interpolant acts as a filtered version of the execution log, explaining the violation of $\phi$. 

Given the information received from \KMReasoner, \KMEAdvisor postulates two alternative approaches to resolve the situation: either to avoid $I$, or to weaken property $FR_1$. The first approach corresponds to restricting the user's ability to execute $cancel\_override$ when such execution would lead to violating property $FR_1$. The second approach corresponds to relaxing the system requirements so that they cease to be inconsistent with the observed system execution.

\paragraph{Exploring the avoidance of the conflicting situation}
\KMEAdvisor explores whether the violation of property $FR_1$ could be avoided by strengthening the constraints on the execution of the $cancel\_override$ transition in Figure~\ref{fig:updated_sm}. As a first exploration, the agent observes that adding $temp \neq 28$ as a guard on $cancel\_override$ would straightforwardly prevent the specific violation observed. However, upon further analyzing the feasibility of this option, \KMEAdvisor notes that such guidance would be too narrow: it would still allow violations of $FR_1$ when the transition is taken with, for example, $temp = 29$. 

To assess whether a more general solution exists under this same strategy, the advisor instructs \KMReasoner to compute the weakest precondition of the $cancel\_override$ transition that guarantees property $FR_1$. The computed weakest precondition is $20 \leq temp \leq 26$, which would provide a more comprehensive and reusable condition. Based on this analysis, \KMEAdvisor formulates the following potential guidance for the modeling subsystem, corresponding to this violation avoidance approach: 
\begin{quotation}
\small
\noindent\textsc{Strengthening Guard Guidance:}\par
\vspace{0.3em}
    Strengthen the guard of transition $cancel\_override$ with constraint $20 \leq temp \leq 26$.
\end{quotation}
This guidance would be both feasible—respecting requirement $FR_1$—and general, covering all temperature values that would otherwise lead to a violation. \KMEAdvisor thus retains this as a candidate strategy, pending further evaluation against alternative approaches.

\paragraph{Exploring requirements weakening}

An alternative approach to altering the transitions in the KM is to explore whether the formalization of the requirements is inappropriate or too restrictive with respect to the requirements engineers' intent. \KMEAdvisor examines this possibility by employing LLM-based probabilistic inference to assess whether a modification of requirement $FR_1$ could resolve the observed inconsistency. 

As part of this exploration, \KMEAdvisor considers relaxing $FR_1$ to give the system time—when in automated mode—to eventually reach the desired temperature. Under this relaxation, the observed execution log would no longer contradict the requirement. The advisor synthesizes the following parametric STL formalization as a candidate replacement for $FR_1$:
$$\Box \big((\Box_{[0,cd]} Auto\_Control) \implies \Diamond_{[cd,cd]} 20 \leq temp \leq 26\big).$$
This formula states that if the system has continuously remained in auto mode for $cd$ units of time, then the temperature must have reached a value within the 20–26°C range by that time. Of course, variable $cd$ must be instantiated with an appropriate time value. To explore feasibility, \KMEAdvisor requests \KMReasoner to synthesize such a value, which performs dynamic specification inference by observing system logs and infers a suitable value for $cd$, e.g., 20 minutes.

Based on this analysis, \KMEAdvisor formulates the following potential guidance for the modeling subsystem, corresponding to this requirements relaxation approach:
\begin{quotation}
\small
\noindent\textsc{Relaxing Requirement Guidance:}\par
\vspace{0.3em}
    Relax the original requirement to allow the system a bounded grace period after entering auto mode, during which the temperature may temporarily fall outside the desired comfort range, provided the target range is achieved within that period.
\end{quotation}
The advisor thus retains this as a second candidate strategy, pending a final decision on which of the two approaches—strengthening transition guards or relaxing the requirement—is more appropriate given additional contextual information and system goals.

\paragraph{Selecting guidance from alternative scenarios}
Having explored both candidate strategies, \KMEAdvisor now integrates additional information to arrive at a final decision. First, the advisor considers the nature of the requirement itself: $FR_1$ was confirmed, through knowledge retrieval and LLM-based probabilistic inference, to be a correct formalization of a hard requirement intended to ensure occupant comfort and safety. Strengthening the guard on $cancel\_override$ would enforce compliance with $FR_1$ but could also prevent legitimate user overrides in borderline cases (e.g., when the temperature is 27°C), potentially degrading user experience. Second, the advisor examines historical operational data from \textit{Execution Base}, which indicates that the system typically reaches the target temperature range within 15–20 minutes under normal conditions. Third, the advisor considers applicable standards and domain knowledge, which suggest that temporary deviations from the comfort range are acceptable as long as they are resolved within a reasonable time frame.

Based on this integrated analysis, the advisor determines that relaxing the requirement, rather than restricting user actions, is the more appropriate strategy, as it preserves user autonomy while maintaining safety and comfort over time. Consequently, the guidance provided to the modeling subsystem is as follows:
\begin{quotation}
\small
\noindent\textsc{Weaken Requirements Guidance:}\par
\vspace{0.3em}
    Weaken the requirement that the temperature must be in the 20–26°C range when in auto mode, to allow a period of up to 20 minutes to achieve that range.
\end{quotation}

\subsection{Empowering MBT}
\textit{TrustModel} can support diverse MBT activities, ranging from test-model construction and evolution to test generation, test execution, integration and system testing, regression testing, and coverage analysis.
Therefore, in the MBT context, the agents, including \KMConstructor, \KMEvolver, \KMEAdvisor and \KMExecutionDriver, can be developed or adapted to address test objectives and testing activities, rather than being used only for general KM construction and evolution.
Table~\ref{tab:MBT} summarizes the main responsibility, inputs and outputs of each component of TrustModel in the context of MBT.

In the testing context, the KM represents test models, i.e., the expected behavior of the SUT, and serves as the basis for deriving tests, executing them, and evaluating observed behaviors against expected behaviors.
In the modeling agentic subsystem, \KMConstructor and \KMEvolver can be used to construct test models from requirements, design artifacts, and testing goals.
These models capture the expected behavior of the SUT and provide a structured basis for subsequent test generation and execution.

In the evolution agentic subsystem, \KMEAdvisor can incorporate relevant domain knowledge to assess the KM from a semantic perspective.
It can also generate \textit{guidance} that specifies expected behaviors, testing assumptions, and potential test oracles, thereby supporting KM refinement and evolution.

In the conformance agentic subsystem, \KMExecutionDriver can work with \KMExecutor and \KMValidator to generate triggers for achieving different testing objectives. 
At the early stage, these triggers can be used to explore the behavior space of the SUT (e.g., reward with new behavior discovered, code coverage).
In later stages, the refined KM can be exploited to generate targeted triggers for specific objectives, such as regression testing, coverage improvement, fault detection, and system-level validation.
Based on test model execution results and the resulting system data, \KMEAdvisor can analyze the feedback to refine the KM and update subsequent testing guidance.
Together with \KMExecutor and \KMValidator, the KM can support test execution and generate test reports with behavioral coverage information, while also serving as a basis for deriving concrete test cases.

\textit{TrustModel} therefore enables an MBT-oriented workflow in which test models are constructed, knowledge-guided, executed, and continuously refined according to testing objectives.

\begin{table}[htbp]
\small
\centering
\caption{Summary of Responsibilities, Inputs and Outputs of Each Component of \textit{TrustModel} for MBT}
\label{tab:trustmodel-mbt}
\resizebox{.99\textwidth}{!}{
\renewcommand{\arraystretch}{1.35}
\begin{tabular}{p{3.5cm} p{5.5cm} p{5.9cm}}
\hline
\textbf{Subsystem/Component} 
& \textbf{Responsibility} 
& \textbf{Input and Output} \\
\hline

\rowcolor{gray!20}
\textbf{Modeling}
& Constructs, updates, and analyzes test models representing expected SUT behavior.
& \textbf{Input:} Test basis, metamodels, guidance \newline
  \textbf{Output:} test models
\\
\hline

\textit{Model Constructor \newline (Agent)}
& Constructs initial test models from test basis and testing goals.
& \textbf{Input:} Test basis, e.g., requirements, design artefacts; testing goals; metamodels, e.g., SysML v2 \newline
  \textbf{Output:} Initial test models
\\
\hline

\textit{Model Updater \newline (Agent)}
& Refines existing test models according to system changes and guidance.
& \textbf{Input:} Test basis, metamodels, guidance, existing test models, i.e., TestModel\textsubscript{pre} \newline
  \textbf{Output:} Updated test models
\\
\hline

\textit{Model Analyzer}
& Checks test models for syntactic, structural, and diagnostic issues.
& \textbf{Input:} Test models \newline
  \textbf{Output:} Analyzed results, e.g., syntax errors, structural inconsistencies, diagnostic findings
\\
\hline

\rowcolor{gray!20}
\textbf{Conformance}
& Executes test models and assesses conformance of observed SUT behavior.
& \textbf{Input:} Test models, testing objectives \newline
  \textbf{Output:} Conformance feedback
\\
\hline

\textit{KM Execution Driver \newline (Agent)}
& Generates test-model-guided execution triggers for MBT activities.
& \textbf{Input:} Test models, execution results, validation results, testing objectives \newline
  \textbf{Output:} Execution triggers, e.g., system events, user actions, environment stimuli
\\
\hline

\textit{KM Executor}
& Executes triggers on the SUT and its operating environment.
& \textbf{Input:} Test models, triggers, API schemas to the SUT and operating environment \newline
  \textbf{Output:} Execution results, e.g., execution traces, system responses, logs, runtime data
\\
\hline

\textit{Runtime Validator}
& Checks execution results against expected test-model behavior.
& \textbf{Input:} Test models, execution results \newline
  \textbf{Output:} Validation results, e.g., missing transitions, mismatched target states, violated assumptions, behavioral deviations
\\
\hline

\rowcolor{gray!20}
\textbf{Evolution}
& Guides refinement and evolution of test models.
& \textbf{Input:} Software lifecycle artefacts, e.g., test basis, domain knowledge, conformance feedback \newline
  \textbf{Output:} Guidance for test model refinement and evolution
\\
\hline

\textit{Evolution Advisor \newline (Agent)}
& Provides semantic guidance for refining test models.
& \textbf{Input:} Software lifecycle artefacts, domain knowledge, conformance feedback \newline
  \textbf{Output:} Guidance, e.g., expected behaviors, testing assumptions, potential test oracles, constraints, refinement suggestions
\\
\hline

\textit{Symbolic Reasoner}
& Supports test model evolution decisions through reasoning.
& \textbf{Input:} Test models, conformance feedback, guidance \newline
  \textbf{Output:} Reasoning results, e.g., suggested model changes, impact analysis, evolution decisions
\\
\hline

\end{tabular}

}
\end{table}


\section{Discussions}\label{sec:discussion}
Recall that \textit{TrustModel} is a multi-agent platform for constructing, validating, and evolving living KMs, ensuring that KMs remain continuously synchronized with the software system and its operating environment. Through the collaboration of its three agentic subsystems (i.e., modeling, conformance, and evolution), \textit{TrustModel} continuously maintains trustworthy and up-to-date KMs.

In Section~\ref{sec:TM4MBT}, we demonstrated how \textit{TrustModel} can be used to maintain test models for enabling MBT. More generally, the continuously maintained KMs can serve as foundational assets for a wide range of MDE activities. In the remainder of this section, we discuss several potential applications of \textit{TrustModel} beyond MBT, and summarize our understanding of how \textit{TrustModel} should be utilized. 

\subsection{Requirements and Assumption Monitoring}  
To perform this MDE activity, the three subsystem architecture is motivated by the need to continuously assess whether requirements and assumptions captured in a KM remain valid as the system and its operating environment evolve. 

The modeling agentic subsystem maintains a requirements-oriented KM that captures requirements, assumptions, environmental expectations, constraints, and their traceability relationships. This KM serves as the reference representation of what the system is expected to achieve and under which conditions these expectations are assumed to hold.

The conformance agentic subsystem is responsible for collecting evidence from \textit{Execution Base}. To achieve this, a \textit{Monitoring Orchestrator} agent (comparable to the \textit{KM Execution Driver} agent of the \textit{TrustModel} for MBT architecture) is needed to derive observation strategies from the requirements-oriented KM and generates execution triggers that drive interactions with \textit{Execution Base}. This agent can be implemented as an LLM-based agent because its primary responsibility is to interpret requirements and assumptions and transform them into executable observation strategies.
The resulting execution traces and environmental observations are collected and analyzed by \textit{Runtime Validator}, which can rely on symbolic and rule-based mechanisms to ensure trustworthy conformance assessment.

The evolution agentic subsystem is responsible for interpreting detected nonconformances between the KM and reality. Such nonconformances may indicate requirement violations, invalidated assumptions, environmental changes, or obsolete requirements. \textit{Evolution Advisor} (same as for \textit{TrustModel} for MBT) analyzes the runtime conformance feedback, supported by symbolic reasoning over requirements, assumptions, constraints, and traceability links, and generates guidance for updating the KM. In this way, the platform supports not only runtime monitoring but also the continuous evolution of requirements and assumptions.

Together, the three subsystems establish a closed-loop process in which the modeling subsystem defines the expected system behavior and operating assumptions, the conformance subsystem gathers evidence from the running system and its environment, and the evolution subsystem updates the KM when expectations and reality diverge. This enables continuous requirements and assumption monitoring in the presence of evolving systems and changing operational environments.

\subsection{Architectural Drift Tracking}
As software systems become increasingly complex, long-lived, and collaboratively developed by multiple teams, maintaining an accurate architectural representation becomes increasingly difficult. Architectural models are often created during design but are rarely updated at the same pace as the implementation. Consequently, architectural drift, i.e., the gradual divergence between the intended architecture and the actual implementation, is commonly observed in practice. Such drift can reduce the usefulness of architectural knowledge for maintenance, impact analysis, and system evolution. \textit{TrustModel} can support architectural drift tracking by maintaining an architecture-oriented KM as a living representation of the intended system architecture.

In the modeling subsystem, the \textit{Model Constructor} agent builds an architectural KM from architecture descriptions, design documents, source code artifacts, dependency information, and other available architectural knowledge sources. The architectural KM captures not only architectural structures, such as components, interfaces, connectors, dependencies, and deployment relationships, but also \textit{architectural intent}, including architectural decisions, assumptions, constraints, design rationale, and quality-attribute objectives (e.g., all external requests must pass through an API gateway). As the architecture evolves, the \textit{Model Updater} agent incorporates approved architectural changes into the KM. Throughout this process, both agents interact with \textit{Model Analyzer} to validate the consistency, completeness, and structural correctness of the architectural KM.

The conformance subsystem is responsible for detecting architectural drift by assessing the consistency between the architectural KM and the implemented architecture. 
The \textit{Architectural Drift Assessor} agent derives design decisions and observation strategies from the architectural KM and periodically generates execution triggers.
These execution triggers specify not only the system inputs that drive active execution of the software system within \textit{Execution Base}, such as invoking service endpoints, replaying representative user requests, injecting test events, or executing workload scenarios, but also checks to be performed through passive scanning, such as analyzing technology stack usage, architectural rules, dependency graphs, or configuration compliance. 
\textit{KM Executor} collects architectural evidence, including runtime observations such as service invocations, message exchanges, interaction traces, and workload responses, and structural observations such as dependency relations, configuration states, and interface specifications.
\textit{Runtime Validator} reconstructs an observed architectural view from the collected evidence and evaluates it against the architectural structures, constraints, assumptions, and design decisions encoded in the KM.
Architectural drift is identified when discrepancies are detected between the intended and observed architectures, such as undocumented dependencies or violations of architectural constraints. The resulting runtime conformance feedback provides evidence for subsequent analysis and model evolution.

Similar for supporting MBT and other MDE activities, the evolution subsystem consumes the architectural conformance feedback and reasons about detected discrepancies using the architectural knowledge captured in the KM. The \textit{Evolution Advisor} agent then generates guidance for updating architectural elements, assumptions, constraints, and design decisions, thereby keeping the architectural KM synchronized with the evolving implementation and its architectural intent. 

\subsection{Change Impact Assessment}
Software systems continuously evolve to accommodate new requirements, fix defects, improve quality attributes, and adapt to changing environments. Understanding the consequences of a proposed change before it is implemented is therefore essential for effective system maintenance and evolution. However, change impacts often propagate across requirements, architecture, implementation, operational assumptions, and environmental conditions, making them difficult to assess using individual artifacts alone. \textit{TrustModel} addresses this challenge by maintaining a living KM that continuously captures and evolves such cross-artifact relationships.

In this scenario, the modeling and evolution subsystems play the central role, while the conformance subsystem is not necessarily required. This is because change impact assessment does not always require driving the execution of the software system or comparing runtime behavior against the KM. Instead, the change impact KM must be continuously updated as connected engineering artifacts evolve. The modeling subsystem constructs and maintains the KM from requirements, architectural models, implementation artifacts, interface specifications, deployment configurations, assumptions, and traceability links. The evolution subsystem analyzes changes in these connected artifacts and generates guidance for updating affected dependencies, links, assumptions, and relationships in the KM. Impact assessment can then be performed by reasoning over the maintained KM to estimate the consequences of a proposed change across related artifacts and system concerns.

\subsection{Summary}
The above discussions show that the high-level architecture of \textit{TrustModel}, organized around the three agentic subsystems of modeling, conformance, and evolution, provides a general foundation for supporting different MDE activities. In particular, the detailed MBT instantiation demonstrates how this architecture can be specialized into a concrete solution in which a test-oriented KM, i.e., test model, is constructed, executed together with the software system and its operating environment, validated through conformance assessment, and evolved based on the resulting feedback. Similarly, requirements and assumption monitoring and architectural drift tracking can be supported by instantiating \textit{TrustModel} with different types of KMs and task-specific conformance criteria.

However, \textit{TrustModel} should not be viewed as a universal replacement for all MDE techniques. When applying \textit{TrustModel} to a new MDE activity, the first question is whether the activity critically depends on keeping a KM alive, that is, continuously synchronized with the evolving system, its engineering artifacts, and its operating environment. If the activity can be sufficiently supported by a static model, a one-time analysis, or artifact-specific techniques, then the full \textit{TrustModel} architecture may not be necessary. In such cases, only selected mechanisms, such as KM evolution or conformance checking, may be useful.

If maintaining a living KM is essential, the next question is how KM evolution should be triggered. For execution-centered activities such as MBT, and in some cases requirements monitoring or architectural drift tracking, KM evolution may be driven by executing the KM together with the software system and its operating environment, namely Execution Base. In this case, the conformance subsystem plays a central role: it generates or coordinates execution triggers, collects evidence from Execution Base, validates the observed behavior against the KM, and produces conformance feedback for the evolution subsystem.

In contrast, not all MDE activities require execution-based conformance assessment. As discussed for change impact assessment, the main requirement is to maintain an up-to-date KM that captures cross-artifact dependencies and traceability relationships among requirements, architecture, implementation, interfaces, deployment configurations, assumptions, and environmental factors. In such cases, KM evolution may be triggered by changes in connected engineering artifacts rather than by executing the software system. The conformance subsystem is therefore not always necessary; the modeling and evolution subsystems may be sufficient to maintain the KM and support reasoning over it.

Overall, \textit{TrustModel} is most suitable for MDE activities where the value of the analysis depends on the freshness, consistency, and trustworthiness of the underlying KM. The framework provides a reusable architectural pattern, but each concrete instantiation should be designed by identifying the required KM type, the source of KM evolution triggers, the need for execution-based conformance assessment, and the form of feedback required to keep the KM alive. In practice, developing such instantiations also requires the effective combination of existing technologies and tools, such as modeling environments, traceability management tools, runtime monitoring or logging infrastructures, static and dynamic analysis techniques, rule engines, symbolic reasoners, and LLM-based agents. \textit{TrustModel} is therefore not intended to replace these technologies, but to organize and integrate them around the lifecycle of living KMs.

As more \textit{TrustModel} applications emerge, we also see an opportunity to abstract reusable modeling, conformance, and evolution subsystems. Such generic subsystems could provide common services for KM construction and update, model analysis, execution or observation coordination, conformance assessment, feedback interpretation, and KM evolution guidance. This would facilitate the rapid development of \textit{TrustModel} architecture instances for different MDE tasks, where developers can specialize the KM type, conformance criteria, evolution triggers, and supporting tools rather than redesigning the whole architecture from scratch.






\section{Conclusion and Outlook}\label{sec:conclusion}
Complex software systems are increasingly expected to operate, adapt, and remain dependable in environments that are dynamic, uncertain, and continuously evolving. In such settings, knowledge models (KMs) are not merely documentation artifacts, but essential engineering assets for making system knowledge explicit, traceable, and actionable. They provide a basis for reasoning about system behavior, assumptions, constraints, design decisions, and operating contexts. As systems evolve, however, these models may themselves become incomplete, inconsistent, or outdated. Maintaining their trustworthiness is therefore a prerequisite for dependable engineering.

This paper introduced \textit{\textit{TrustModel}}, a vision for the agentic generation and evolution of living KMs. By organizing this vision around three agentic subsystems: \textit{Modeling}, \textit{Conformance}, and \textit{Evolution}, \textit{TrustModel} provides a conceptual foundation for creating and continuously evolving KMs. The modeling subsystem supports the creation and update of KMs; the conformance subsystem evaluates their alignment with the actual system and its operating context; and the evolution subsystem generates guidance for resolving inconsistencies, incompleteness, and obsolescence.

Through its instantiation for MBT, \textit{TrustModel} illustrates how living KMs, i.e., test models, can support more adaptive and MBT practices. At the same time, the vision extends beyond testing to broader model-driven engineering (MDE) activities, including requirements and assumption monitoring, architectural drift tracking, and change impact assessment. These applications suggest that living KMs can serve as a unifying basis for maintaining engineering knowledge across the software system lifecycle. In particular, for constantly evolving complex software systems, building and sustaining such KMs is essential: engineering assumptions need to remain transparent, system and environmental changes need to be continuously reflected, and decision-making processes need to be evaluated against explicit models.

More broadly, we expect this paper to contribute to a renewed discussion on the role of MDE and model-based systems engineering (MBSE) in the engineering of modern complex software systems. The long-standing ambition of MDE has been to raise the level of abstraction, make engineering knowledge explicit, and enable automation based on models. 
Recent advances in large language models (LLMs) and agentic AI provide a timely opportunity to reshape how such automation can be realized, especially for the continuous evolution of models together with the systems and operating environments they represent. Rather than replacing models, LLM-based agents may help MDE and MBSE overcome some of their long-standing barriers to automation. Traditionally, model-based automation has depended heavily on carefully specified transformation rules across engineering artifacts, manually constructed metamodels, and dedicated tools that offer only limited support for interpreting natural-language requirements and other informal sources of knowledge. As a result, building KMs from heterogeneous artifacts has remained effort-intensive, and evolving them has been a persistent challenge, whether performed manually or through automated mechanisms. With LLMs, and with carefully selected tools integrated into agentic workflows, we now see an opportunity to more readily connect heterogeneous artifacts, derive semantics from them, and support the continuous creation, checking, evolution, and operationalization of KMs at a scale and pace that were previously difficult to achieve.

Future work will focus on realizing \textit{TrustModel} for MBT in the context of ModelCopilot\footnote{ModelCopilot: https://www.modelcopilot.org/model-copilot.html}. We will also investigate how \textit{TrustModel} can be instantiated for other MDE activities, such as requirements and assumption monitoring, architectural drift tracking, and change impact assessment. In addition, we plan to evaluate the \textit{TrustModel} architecture and its concrete instantiations in real-world software engineering contexts, with particular attention to their effectiveness in maintaining trustworthy, living KMs under continuous system and environmental change. In the long term, \textit{TrustModel} aims to support dependable engineering practices in which KMs continuously evolve with the systems they describe. We hope that this vision will call for further contributions from the MDE, MBSE, software engineering, and AI communities toward trustworthy, living KMs for continuously evolving software-intensive systems.

\bibliographystyle{ACM-Reference-Format}
\bibliography{bib}

\end{document}